**Recommended citation**

Atoum, I., & Bong, C. H. (2015). Measuring Software Quality in Use: State-of-the-Art and Research Challenges. *ASQ.Software Quality Professional*, *17*(2), 4–15.

URL http://asq.org/software-quality/2015/03/implementation-and-certification-of-iso-iec-29110-in-an-it-startup-in-peru.pdf

http://asq.org/pub/sqp/past/vol17_issue2/index.html

@article{Qinu,
author = {Atoum, Issa and Bong, Chih How},
institution = {ASQ},
journal = {ASQ.Software Quality Professional},
number = {2},
pages = {4--15},
title = {{Measuring Software Quality in Use: State-of-the-Art and Research Challenges}},
volume = {17},
year = {2015}
}

# Measuring Software Quality in Use: State-of-the-Art and Research Challenges

**ISSA ATOUM AND CHIH HOW BONG**
Universiti Malaysia Sarawak

Software quality in use comprises quality from the user's perspective. It has gained its importance in e-government applica-tions, mobile-based applications, embedded systems, and even business process development. Users' decisions on software acquisitions are often ad hoc or based on preference due to difficulty in quantita-tively measuring software quality in use. But, why is quality-in-use measurement difficult? Although there are many software quality models, to the authors' knowledge no works survey the challenges related to software quality-in-use measurement. This article has two main contributions: 1) it identifies and explains major issues and challenges in measuring software quality in use in the context of the ISO SQuaRE series and related software quality models and highlights open research areas; and 2) it sheds light on a research direction that can be used to predict software quality in use. In short, the quality-in-use measurement issues are related to the complexity of the current standard models and the limitations and incompleteness of the customized software quality models. A sentiment analysis of software reviews is proposed to deal with these issues.

## INTRODUCTION

With a large amount of software published online it is essential for users to find the software that matches their stated or implied needs (quality in use). Users often seek adequate software quality. It is important to quantify software quality from a user's perspective in order to compare different types of software, thus allowing users to acquire the *best* software quality. To understand what is meant by inadequate or adequate software quality, it helps to define quality. Defining quality is not easy in that quality is based on many possible disciplines: philosophy, economics, marketing, and so on. Garvin (1984) identified five views/approaches of quality. The closest defini-tion to this work is the user-based approach definition "meeting customer needs." As referred to by Deming (2000), Shewhart (1931) defines quality: "There are two common aspects of quality: One of them has to do with the consideration of the quality of a thing as an objective reality independent of the existence of man. The other has to do with what we think, feel or sense as a result of the objective reality. That is to say, there is a subjective side of quality." This side is yet another meaning of *quality in use*. If the customer is satisfied, then a product or service has adequate quality.

# Measuring Software Quality in Use: State-of-the-Art and Research Challenges

Quality in use or perceived quality by users is very important to many applications. Quality in use is gaining more attention, especially in applications where end users are the core of a successful project.
It has been adopted in mobile-based applications (Alnanih, Ormandjieva, and Radhakrishnan 2014; La and Kim 2013; Osman and Osman 2013), Web applica-tions (González et al. 2012; Orehovački, Granić, and Kermek 2013; Orehovački 2011), healthcare applications
(Alnanih, Ormandjieva, and Radhakrishnan 2013; Alnanih, Ormandjieva, and Radhakrishnan 2014), project management tools (Oliveira, Tereso, and Machado 2014), and business process management (Heinrich, Kappe, and Paech 2011; Heinrich 2014).

There are many benefits that can be gained if qual-ity is adopted early in the software development life cycle. Quality in use, a user-centered approach, will get users' feedback early; thus, different ways to access software, including content and user interface, are adopted early, hence software success. A good software design will let the user work effectively and efficiently, saving time; thus, it will increase user productivity.
Moreover, if the user interface (one aspect of quality in use) takes into account user needs and expectations, it will reduce system errors. On the other hand, if the system interface is poorly designed, it will result in increased training time and malfunction errors. On one hand, given these benefits, it is more likely that the user will use the application; on the other hand, software vendors will get improved user acceptance. In fact, systematic evaluation of quality in use is important because it allows continuous software recommendations and improvements.

Moving from the general concept of quality of soft-ware, software quality can be conceptualized from three dimensions: software quality requirements, the quality model, and quality characteristics. Quality require-ments are what the user needs in the software, such as performance, user interface, or security requirements. The quality model is how characteristics are related to each other and to final product quality. Measuring software quality will check if user requirements are met and determine the degree of quality. The final objective of software quality is to know whether the software is of adequate or inadequate quality. To identify the so-called adequate or inadequate quality, a software quality model is needed. The quality model categorizes quality into characteristics overseen by measurement methods. The quality model is a "defined set of characteristics and of relationships between them, which provides a framework for specifying quality requirements and evaluating quality" (ISO/IEC 2005, 7). The quality model identifies the adequate or inadequate quality attributes. The quality characteristic is the "category of software quality attributes that bears on software quality" (ISO/ IEC 2005, 9), simply the main factors or properties that determine whether quality is adequate or inadequate. Measurements are "sets of operations having the object of determining a value of a measure" (ISO/IEC 2011, 52). That is to say, the measurement is the actual score of adequate and inadequate quality attributes.

The ISO/IEC 25010:2010 standard (referred to as ISO 25010 hereafter), a part of a series known as software quality requirements and evaluation (SQuaRE), has two major dimensions: quality in use (QinU) and product quality. The former specifies characteristics related to the human interaction with the system and the latter specifies characteristics intrinsic to the product. QinU is defined as the "capability of a software product to influence users' effectiveness, productivity, safety, and satisfaction to satisfy their actual needs when using the software product to achieve their goals in a specified context of use" (ISO/IEC 2005, 17). The QinU model consists of five characteristics: effectiveness, efficiency, satisfaction, freedom from risk, and context coverage. Figure 1 illustrates the definition of these characteristics.

**FIGURE 1** Definitions of quality-in-use characteristics as defined by the ISO 25010 standard

| Characteristic | Definition |
| --- | --- |
| Effectiveness | Accuracy and completeness with which users achieve specified goals (ISO 9241-11). |
| Efficiency | Resources expended in relation to the accuracy and completeness with which users achieve goals (ISO 1998). |
| Freedom from risk | Degree to which a product or system mitigates the potential risk to economic status, human life, health, or the environment. |
| Satisfaction | Degree to which user needs are satisfied when a product or system is used in a specified context of use. |
| Context coverage | Degree to which a product or system can be used with effectiveness, efficiency, freedom from risk, and satisfaction in both specified contexts of use and in contexts beyond those initially explicitly identified. |

## Problem Statement

This article investigates these problems:

1. There are a limited number of literature reviews on software QinU. Although there are many research articles about software quality, to the authors' knowledge this is the first work that specifically identifies the problems of measuring QinU.
2. There is insufficient research on other possible research directions to tackle the first problem. To the authors' knowledge, few works target to resolve the QinU problem systematically (Leopairote, Surarerks, and Prompoon 2012).

## Research Contributions

This article makes the following contributions:

- Identifies and explains several problems while measuring software QinU using the standard and customized quality models. This article is the first to survey several quality models and explain various challenges to measuring QinU. Most of the challenges related to ISO standard models are deemed to complication and incompleteness of the documents. On the other hand, customized quality models are subject to incomplete models that are designed for their own specific needs.
- Presents a research direction to envision software QinU from software reviews. Given the issues related to measurement, sentiment analysis, an emerging branch of natural language processing, can be used to analyze textual user judgments about software (Mei et al. 2007; Taboada et al. 2011).

# SOFTWARE QUALITY-IN-USE MODELS

There have been many works in software quality models, but to the authors' knowledge no research has been conducted to summarize the main problems of measuring quality in use. Measuring software quality in use can be divided in two main frameworks: standard model frameworks and customized model frameworks.

## Standard Frameworks

There have been many standards that can support software quality, but they do not contain any specific model for software quality, but rather checklist guides. For example, it is suggested in the literature that the ISO 9000 family of standards not be used for software quality (Stelzer, Mellis, and Herzwurm 1997). IEEE Std. 730 supports quality assurance plans. Capability Maturity Model Integration (CMMI) (Ahern, Clouse, and Turner 2008) is a process improvement training and certification program. These standards are not designed to address quality in use or specific characteristics of software product quality.

Recently, the SQuaRE ISO standard series is a result of blending the ISO/IEC 9126 and ISO/IEC 14598 series of standards. The purpose of the SQuaRE series of standards is to assist in developing and acquiring software products with the specification of quality requirements and evaluation. From the stakeholder's viewpoint, the quality requirements are specified, and the quality of the product is evaluated based on this specification using the chosen quality model, quality measurement, and quality management process.

To measure QinU, five divisions of the SQuaRE series must be considered. Figure 2 illustrates the organization of the SQuaRE series representing families of standards, which are further called divisions. To measure QinU effectively, the Quality Management Division has to be considered by taking standard definitions; the QinU model is part of the Quality Model Division; the Quality Measurement Division contains the measurement for-mulas. The Quality Requirement and Quality Evaluation Divisions are used to specify quality requirements and evaluate quality. Thus, the quality model is defined, managed, and measured according to quality require-ments and quality evaluation processes.

**FIGURE 2** Organization of SQuaRE series of international standards

| 2503n<br>Quality requirement division | 2501n<br>Quality model division | 2504n<br>Quality evaluation division |
|---|---|---|
| | 2500n<br>Quality management division | |
| | 2502n<br>Quality measurement division | |
| ISO/IEC 25050 – 25099 SQuaRE extension division | | |

Technically, and more precisely, the QinU Measurement Standard ISO 25022 has to be considered in the context of four other standards: the Measurement Reference Model and Guide ISO 25020; the Measurement of Data Quality 25024, the Measurement of System and Software Product Quality ISO 25023, and Quality Measure Elements Standards ISO 25021. Figure 3 depicts the relationship between the ISO/IEC 25022 and other ISO/IEC 2502n division of standards. The overall picture reveals that many other standards are related. These include ISO 9241-11:1998 Guidance on usability, ISO 9241-210:2008 Human-centered design process for interactive systems, ISO/IEC 25063:2014 Context of use description, and ISO/IEC 25064:2013 User needs report.

It is important to note that product dimension has a direct impact on QinU measurement. For example, the functional suitability, performance efficiency, usability, reliability, and security of product quality will have a significant influence on quality in use for primary users. This will add more complication to the measurement process.

Based on the ISO standards, measuring the QinU is not achieved directly. While the standards provide avenues for customization, they need careful quality assurance to provide apparent integration between related standards. Even more, customizing such standards is not defined clearly in the standards. More issues on these standards can be found in a later section.

Next, other models of QinU, the customized models, are discussed.

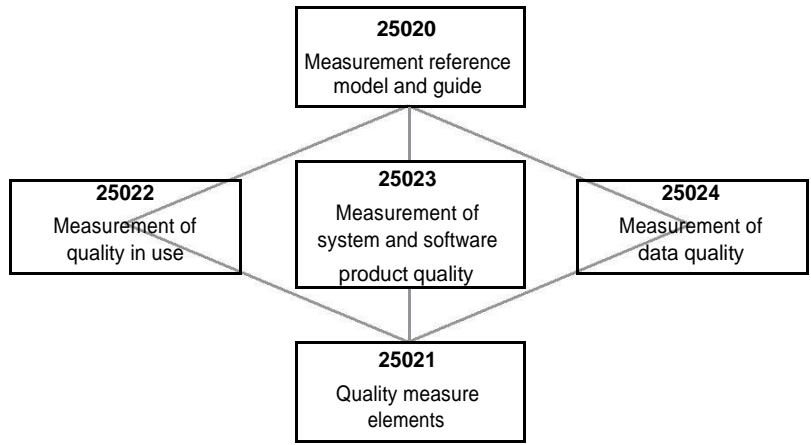

FIGURE 3  Structure of the quality measurement division

# Customized Models

There have been many models that have been developed to measure software quality; some of these models have some of the quality-in-use characteristics such as user experience and usability, but do not consider all the characteristics as defined in the ISO 25010 standard. Moreover, term definitions across different models are sometimes confusing, which limits the accurate mapping of quality-in-use characteristics (García et al. 2006). Following are some related models assembled in logical order. These groups could be overlapping or intersecting with each other.

# Quality-in-Use Specific Models

These models are designed as complete quality-in-use models.

**ISO-based models** These models are based on ISO standard frameworks. La and Kim (2013) adopted the effectiveness and productivity of the ISO 9126 model for a service-based mobile ecosystem. Motivated by the 9126 model structure, Osman and Osman (2013) proposed three categories to model quality in use for mobile government systems. They are usability, acceptance, and user experience. With a defined set of tasks and a post-test questionnaire, they calculated the quality in use. Oliveira, Tereso, and Machado (2014) linked the ISO 9126 characteristics functionality, usability, and operability to a set of criteria/requirement of project management tools in order to help project managers evaluate project management tools. The quality, quality in use, usability, and user experience (2Q2U) model (Lew, Olsina, and Zhang 2010) extends the ISO 25010 model with information quality and learnability in use characteristics, and actual usability and user experience. The Semantic Web Exploration Tools Quality in Use Model (SWET-QUM) (González et al. 2012) extends ISO 25010 with several metrics related to semantic Web exploration tools. In SWET-QUM tasks are customized to Web exploration tools, and are manually linked to the ISO QinU model.

**Strategic quality models** Quality in use can be achieved indirectly by affecting product quality, using a new evaluation technique, or by using a completely new framework. Becker, Lew, and Olsina (2012) proposed SIQinU (a strategy for understanding and improving quality in use) to evaluate product quality in order to recognize quality-in-use problems. They also suggested improving quality in use by changing product quality attributes' operability (learnability, ease of use) and information quality. Alaa, Menshawi, and Saeed (2013) employ the addition of coupling, data complexity, and abstraction and file transfer properties as product quality attributes to enhance quality of service and quality in use. Six quality-in-use metrics models are proposed (Orehovački, Granić, and Kermek 2013; Orehovački 2011) based on literature metrics and new metrics suggested by the work. The metrics are: system quality, service quality, content quality, performance, effort, and acceptability. The authors used an eye-tracking technique (retrospective think aloud) and a questionnaire to measure quality in use. They apply the model on mind mapping and diagram Web 2.0 applications.

In a similar work, Hsu and Lee (2014) proposed using a decision-making method called decision-making trial and evaluation laboratory (DEMATEL) to evaluate the quality of blogging interface. DEMATEL analysis identified the causal relationships between eight blog interface factors using the impact-relations map. Ardito et al. (2014) proposed the pattern-based inspecting method exploiting a list of quality-in-use evaluation patterns to evaluate e-learning systems. In their model, usability problems are linked to platform graphical design, feedback, naviga-tion, and functionality. The study showed that in some cases the pattern-based technique is time demanding.

Quality in use is proposed as an added value to business process development (Heinrich et al. 2011). The added measures are extracted from the literature. Heinrich, Kappe, and Paech (2014) proposed the Business Process Quality Reference Model to provide a comprehensive understanding of business process quality.

**Paradigm-specific models** Some quality models are specific to software paradigm style (procedural vs. object-oriented approach) or programming language. To ease and manage the design of object-oriented software, the strategy factor model is proposed (Marinescu and Ratiu 2004). It explicitly relates the quality of a design to its conformance to a set of essential principles, rules, and heuristics. Bansiya and Davis (2002) adapted reusability, flexibility, understandability, functionality, extendibility, and effectiveness from ISO 9126 and designed 11 proper-ties of object-oriented languages. The Quamoco Product Quality Model (Wagner et al. 2012) harvested 200 factors and 600 measures specialized for Java and C# systems. The Software Quality in Development (SQUID) approach (Kitchenham et al. 1997) is a composite model based on ISO 9126. It defines a structure of model elements and their interactions and set of linked entities using this structure. SQUID is designed for the telescience project and is more conserved with product quality rather the QinU.

## Hierarchical Models

These models link various quality characteristics together at different levels, which in turn are finally linked with root product quality. Following are some examples of these quality models.

**McCall's quality model** (McCall, Richards, and Walters 1977) This model was developed for the U.S. Air Force and is primarily aimed toward system developers and the system development process. It attempts to bridge the gap between users and developers by adopting quality factors that affect the user's and developer's views. The product operation major perspective of the McCall model can be mapped to the QinU; however, it is not complete.

**Boehm's quality model** (Boehm et al. 1978) Boehm's model—similar to the McCall quality model—attempts to qualitatively define a hierarchical quality software model. The hierarchical quality model is structured around high-to intermediate-level characteristics that contribute to the overall quality level. The high-level characteristics address as-is utility, maintainability, and portability. As-is utility can be mapped to the QinU of the ISO standard, but it is not complete since it omits the risk mitigation characteristic.

**Dromey's quality model** (Dromey 1995) Dromey states that quality characteristics cannot be built directly into the software; the quality evaluation differs for each product; and a dynamic process is needed to cover dif-ferent systems. Dromey distinguishes between product components from externally visible quality attributes. The QinU of the ISO standard is implied by both the descrip-tive and correctness characteristics of the Dromey model.

## Usability-Based Models

There are many works that aim to measure one aspect of quality in use called *usability*. The International Organization for Standardization defines usability as the effectiveness, efficiency, and satisfaction with which specified users achieve specified goals in particular envi-ronments (ISO 1998). There is no consensus agreement on this definition; it might refer to user interface, ease of use, or user friendliness (Bačíková and Porubän 2014; Carvajal et al. 2013; Holzinger 2005; ISO 1998; Shackel and Richardson 1991). Usability is frequently linked with user experience with the software (Brooks and Hestnes 2010; Masip, Oliva, and Granollers 2011; Scholtz 2006). According to the definition of ISO 25010, usability, or sometimes called friendliness, is not covering freedom from risk or context coverage characteristics. The intended goal of this work is to highlight the problems of quality in use covering the global definition of quality in use as per ISO 25010. Following is a sample of related works to usability.

**Medical domain** Huang and Chiu (2014) proposed a new process to evaluate the usability of the healthcare system (Kincet Game) to enhance the user experience. The process predefines a set of tasks to be followed by users. A questionnaire was then used to evaluate the usability of the system. Bond et al. (2014) proposed to evaluate two types of medical software (electrocardiogram [ECG] viewer) and (electrode misplacement simulator [EMS]), in the expert conference settings using a set of predefined tasks, and by saving user interactions with the systems.

**Knowledge concepts** Horkoff et al. (2014) proposed a model to evaluate the expressiveness and effectiveness of requirement modeling language (Techne) by using a case study on the three requirement problems. Weinerth et al. (2014) highlighted that there is insufficient research on usability over computer-based concept map assessments. They showed that test takers are affected by the quality (usability) of the tested systems.

**E-applications** Huang and Benyoucef (2014) designed a case study for specific e-government websites to test usability and credibility. Usability and credibility were built based on extended guidelines of Fogg et al. (2001) and Nielsen (1999). Alepis and Virvou (2014) interviewed a set of instructors and students to evaluate the effectiveness of a customized mobile-based application. The interviews were basically based on specific questions about the user friendless of the application. Zheng et al. (2014) proposed a quality model for services based on customized criteria (tasks) and using six quality aspects, including usability.

## Provider-Specific Models

Here is a set of specific quality models of certain organizations.

**FURPS quality model** This model was presented by Grady (1992) and later extended and owned by IBM Rational Software (Ambler, Nalbone, and Vizdos 2005; Jacobson 1999; Kruchten 2004). FURPS stands for functionality, usability, reliability, performance, and supportability. The FURPS+ includes additionally design constraints, implementation requirements, interface requirements, and physical requirements. The QinU of the ISO standards is implied by usability and performance characteristics.

**SAP Q-Index** SAP developed and implemented three quality management systems certified according to ISO 9000, namely SAP Global Development, SAP Active Global Support, and SAP IT.

**The Motor Industry Software Reliability Association (MISRA)** MISRA developed MISRA C for C programming language to facilitate code safety, portability, and reli-ability in the context of embedded systems.

# Summary of Studied Quality-in-Use Models

A direct mapping between different quality-in-use models is difficult due to the tight relationship with product quality and discrepancies in used terminologies; however, several conclusions from the previously discussed works are drawn. To enable a fair comparison between quality models, a baseline is needed. Whether to choose custom-ized or standard models is a dilemma. Many quality-in-use models are context specific and there is no complete model. Kläs et al. (2014) proposed using comprehensive quality model landscapes (CQMLs) as a scheme for quality model relationship comparison. Bakota et al. (2011) identified that a good quality model should be interpretable, explicable, consistent, scalable, extendible, and comparable. However, these approaches are subjective. In this article the ISO 25010 standard quality model is the chosen baseline to compare with other models because it is assumed to be an acceptable approach globally. Figure 4 shows the QinU characteristics of ISO 25010 as compared to studied works.

From the previously studied models, several chal-lenges must be tackled. Following is a list of challenges.

## QUALITY-IN-USE CHALLENGES

For readability and understandability, following are some major challenges that can be faced while measuring software quality in use in general, measuring quality in use using standard frameworks, and measuring quality in use using customized models.

## General Challenges

**Task measurement** To measure QinU there is a need to agree with the software user on a set of tasks that he or she needs to execute to accomplish a pragmatic goal ("do goals" to achieve the task, such as pay a bill). This means the user should be involved in the quality requirements specification, which might not be applicable at all times. Another issue related to task measurement embraces the variability of tasks from one software function to another and from one software to another. For example, a task to open a file for writing is different than a task of removing special characters from a text file. Worse off, defining *what the tasks are* is by itself a major challenge. Hedonic tasks (the "be goals") that imply user satisfaction cannot be specified; thus, they cannot be measured directly.

**FIGURE 4** Definitions of quality-in-use characteristics as defined by the ISO 25010 standard

| Characteristic (from ISO 25010) | Conclusions (deduced from QinU models) |
|---|---|
| Effectiveness | Many models suggested adoption of lower properties of effectiveness, such as the user interface (Alnanih, Ormandjieva, and Radhakrishnan 2014; Hsu and Lee 2014), navigation (Ardito et al. 2014) and learnability (Lew, Olsina, and Zhang 2010), while others employ a higher-level usability to represent this characteristic. It is concluded that a consensus agreement should be applied to effectiveness properties. Hence, the literature lacks research to measure the accuracy of achieving a goal to complete the effectiveness definition. |
| Efficiency | Some quality models add the role of users as resources in quality requirement process (Bakota et al. 2011); conversely, the resources can also include the computer system infrastructure that is consumed to achieve user goals. Possibly these resource statistics are collected during run time as product quality model characteristics. So, it is believed a linkage with software reliability, performance, and portability can enhance this characteristic calculation. |
| Freedom from risk | This characteristic tends to be vague in studied works. Many works do not consider the data loss, for example, as a potential risk to software usage. In the ISO 25010, it is not clear how it is linked to the measurement of data quality. ISO 25024 information has been added as part of quality model (Becker, Lew, and Olsina 2012; Lew, Olsina, and Zhang 2010), but it was not shown how it will be used. |
| Satisfaction | The subjective parts of quality in use were modeled by user experience (Lew, Olsina, and Zhang 2010) and learnability. In fact the effectiveness, efficiency, and risk mitigation are direct factors to satisfaction characteristic. Quantifying this characteristic is difficult, as satisfaction can be just word of mouth. |
| Context coverage | This characteristic has not yet been studied intensively in the literature. It has been shown that users have different profiles, software platforms, educational background, and interests. Also, the software itself has different trends in its life cycle. The software may gain attraction at the beginning, but later it may get less attraction due to extra features needed by the users or other service-related factors such as software delivery and support. The assumption that all users should have used the software in its context of use may be a vague assumption. Consequently, this characteristic needs more focus. |

**The Web software development life cycle** Users of publicly available online software are rarely asked to be part of the system development life cycle, but usually the software companies/developers make assumptions on user needs. In cases where software is designed to be used by global users such as operating systems or antivirus software, then software companies have to find other ways to determine user needs. However, it might be a disaster when users start using the software. This is not because of software bugs—that is usually not the major problem—but rather because users are not satisfied. Users need to see software doing what they were *thinking of* without draining their minds with the life cycle of the software.

**Dynamic customer needs** Customer needs are dynamic and they can change from time to time, so quantitative measures might not be suitable. Ishikawa (1985) states that "We must also keep in mind that consumer requirements change from year to year and even frequently updated standards cannot keep the pace with consumer requirements." These needs are usually resolved by new versions of software; however, software might get complicated or buggy due to extra features added that were not planned ahead. If users are involved ahead of time, these needs might be planned for in advance. Therefore, this problem returns one to the first and second unsolved issues mentioned previously.

# Challenges Related to Standard Quality Frameworks

**Quality model critiques** There are problems intrinsic to quality models. Deissenboeck et al. (2009) identified critiques to many software quality models; they are unclear of their purposes, not satisfying users on how to use the quality models, and there is no uniform terminology between different models (García et al. 2006). Masip, Oliva, and Granollers (2011) stated that user experience is implied in ISO 25010 but is not defined. Deissenboeck et al. (2007) showed that the abilities of ISO/IEC 9126 are abstract and hard to measure. Alnanih, Ormandjieva, and Radhakrishnan (2014) showed that the existing standards do not satisfy the requirements for measuring the quality in use of the mobile user interface in the healthcare domain.

Kläs et al. (2014) showed that "no guidance is available on how to select, adapt, define, combine, use, and evolve quality models." So, the standard models do not clarify how to customize or use quality models; thus, the aggregation of the evaluation score is challenging (Mordal et al. 2013).

**Evaluation requirements** Looking into the mathematical formulas for quality in use in ISO 25022 and the proposed methods to measure quality in use, quality managers might find it a hard job. For example, to measure the effectiveness, task completion, task effectiveness, and error frequency must be calculated.

To get such values an evaluator needs to measure user performance, such as the number of tasks completed, total number of tasks, and number of errors made by users. Practically embedding monitoring tasks with operating software is a major failure because it will consume lots of resources, while software maximum performance is one goal of quality. Even if a usability lab is exploited, it has to be equipped with tools such as eye trackers, monitoring key pressing or mouse movement (González et al. 2012). Using such labs is not feasible for a single user if he or she wants to measure the software quality in use alone.

**Quality standards integration requirements** Measurement of quality in use has to be done in line with the ISO standards ISO 2502n to ISO 25024, and in line with the ISO 25010 model (see Figure 2 and Figure 3). Integrating the related quality processes of these models is a problem for quality engineers. The reason behind this problem is the need of experienced engineers given limited information in the standard models on how to customize them, especially for small-sized companies. In an extension to ISO 25010, Lew, Olsina, and Zhang (2010) suggest adding data quality inside ISO 25010 instead of being separate. Their motivation was that defining data quality separately loses emphasis on using information. The ISO 25022 measurement standard suggests many methods, such as business analysis, risk analysis, inspec-tion, and analysis of user performance. A wide range of measuring methods requires an acceptable level of experts in each domain. For a single user, these domains are not achievable. Moreover, software companies/developers will not add additional monitoring functions to their software to keep an acceptable degree of software reliability.

**QinU boundary factors** While a quality-in-use model tries to measure the human computer system interaction, there are many factors that affect quality in use at the same time. They are: the information system, target computer system, target software, target data, usage environment, and user type (primary, secondary, or indirect user). These factors possibly draw a nonlinear dynamic relationship over time.

## Challenges Related to Customized Models

ISO-based models (González et al. 2012; La and Kim 2013; Lew, Olsina, and Zhang 2010; Oliveira et al. 2014; Orehovački et al. 2013; Orehovački 2011; Osman and Osman 2013) are rather incomplete and inherit the problems of the standard models. Many of them are specialized to certain quality applications: mobile-based applications (Alnanih, Ormandjieva, and Radhakrishnan 2014; La and Kim 2013; Osman and Osman 2013) or Web applications (González et al. 2012; Orehovački et al. 2013; Orehovački 2011). The strategic-based models are not yet mature (Alaa, Menshawi, and Saeed 2013; Hsu and Lee 2014) or incomplete (Becker, Lew, and Olsina 2012). The incompleteness is a result of using a subset of quality-in-use characteristics. These models, for example, do not consider risk mitigation nor context coverage characteristics. Paradigm-specific (Marinescu and Ratiu 2004; Wagner et al. 2012) models are programming language specific, thus limited. Hierarchical models (Dromey 1995; McCall, Richards, and Walters 1977) target the software product or process characteristics and do not suit software quality in use or require user involvement (Al-Qutaish 2010; Samadhiya, Wang, and Chen 2010). The usability-based models (Bačíková and Porubän 2014; Bond et al. 2014; Huang and Chiu 2014; Zheng et al. 2014) are domain specific and do not cover the complete definition of quality in use. This usually excludes freedom from risk and context coverage QinU characteristics. Extending such models might end up with something close to the ISO 25010 QinU model.

So what is a possible solution to these challenges? Next, a research direction based on natural language processing of software reviews is presented.

## FUTURE RESEARCH DIRECTION

Modern Internet technology is an invaluable source of business information. For instance, the product reviews on social media sites composed collaboratively by many independent Internet reviewers through social media can help consumers make purchasing decisions and enable enterprises to improve their business strategies. Various studies show that online reviews have real economic value for the products they target (Ghose and Ipeirotis 2011).

Opinion mining or sentiment analysis is an emerging research direction of natural language processing that targets to analyze textual user judgments about products or services (Mei et al. 2007; Taboada et al. 2011). Reviews' text snippets are a good source for user decision making and a goldmine for a product's reputation. Analysis of reviews text manually is obviously hard. The average reader will have difficulty accurately summarizing relevant information and identifying opinions contained in reviews about a product. Moreover, human analysis of textual information is subject to considerable biases resulting from preferences and different understandings of written textual expressions.

Measuring software quality in use automatically from software reviews is challenging due to the anticipated complexity of designing an appropriate approach for a large volume of software reviews. As the number of reviews is increasing dramatically, going through product reviews can be a painful process and is usually laboriously lengthy.
Most of the time, product reviewing can be confusing. For example, comments like "I just don't like this product" and "The product took forever to be here" lack construc-tive expressions, as these comments are not targeted to the product. Thus, there is a limited human capability to produce consistent results. Therefore, opinion mining is needed to identify important reviews and opinions to answer users' queries (Qiu et al. 2009; Zhang, Xu, and Wan 2012).

A major task of opinion mining involves identifying and summarizing sentimental expressions (Duric and Song 2012; Lu, Zhai, and Sundaresan 2009; Zhang and Liu 2011). Current opinion mining is restricted to reveal bipolarity of the reviews: positive and negative (Shein and Nyunt 2010; Turney 2002). In a bipolar approach under a voting mechanism, the qualities of evaluations are affected by imbalanced vote bias. Thus, there is a vital need for effective evaluation of product quality. Although there are studies reported to extend opinion mining to using a scaling system, they suffer the same predicament: determining the polarity of the review. The flaw with the polarity system is that the overall review can be easily tampered with because of the existing opinion spam.

Furthermore, a truthful reviewing process should center on the efficiency and effectiveness of the product to users (Zhang et al. 2010). The more fine-grain works are on feature or aspect-based sentiment analysis where it determines the opinions on the features of the reviewed entity such as a cell phone, tablet, and so on. However, this has the same problems mentioned previously, as it is difficult to identify relevant entities and polarity. In addition, the review may not focus on the entity itself. To the authors' knowledge, little research has been published in software reviews opinion mining. Mining software reviews can save users time and can help them in the software selection process, which is time consuming.

Despite the difficulties of the sentiment analysis approach, it can be used to overcome the issues discussed previously. Sentiment analysis can seemingly work on user reviews without active user involvement. So, future improvements to quality can be directed. Sentiment analysis has been applied in many domains such as movie reviews (Thet, Na, and Khoo 2010), customer reviews of electronic products (Hu and Liu 2004) or net-book computers (Brody and Elhadad 2010), services
(Long, Zhang, and Zhut, 2010), and restaurants (Brody and Elhadad 2010; Ganu, Elhadad, and Marian 2009).

To the authors' knowledge, few works aim to resolve QinU problems systematically. Bond et al. (2014) proposed natural language processing (NLP) techniques to enhance user friendliness of electronic dictionaries. They used syntactic analysis, morphological analysis of inflection, and word formation to improve guidance of text produc-tion and access to lexical data. Leopairote, Surarerks, and Prompoon (2012) proposed a quality-in-use model based on software reviews. They proposed to seed an ontology with tokens extracted manually from the ISO
9126 document. Then the ontology was expanded with antonyms and synonyms using WordNet. Experts were used to map review sentences to the created ontology. Atoum and Bong (2014) proposed a framework to predict software quality in use from user reviews. Their work is based on using latent Dirichlet allocation (LDA) to identify important aspects of software quality in user reviews
(Blei, Ng, and Jordan 2003; Blei and McAuliffe 2010). The framework employs opinion-feature double propagation to expand predefined lists of software quality-in-use features.

# RELATED WORK

To the authors' knowledge little research has been conducted to identify problems in quality-in-use mea-surement. Most of the reviewed works are not rather comprehensive. Hirasawa (2013) identified several

# Measuring Software Quality in Use: State-of-the-Art and Research Challenges

challenges in QinU management while incorporating quality in use for embedded systems' developments in Japan. The identified challenges relate to terminology disparity, system engineer capabilities, and lack of user process. He suggested implementing user requirements, understanding the outcome as a service rather than a product, and managing functional safety as part of quality in use.

Bevan (2001; 1999) identified several barriers to quality-in-use measurement: 1) incomplete requirements, as user requirements are not taken into consideration early; 2) additional cost resulting from adding user activities and software and hardware requirements in the software life cycle; and 3) barriers resulting from the unclear definition of who should use the system and in which circumstances.

# CONCLUSION

Quality in use represents software quality from a user's viewpoint. This article presents the major issues in measuring software quality in use. Quality in use can be measured using the standard SQuaRE series, while many characteristics of software quality in use are scattered in many customized software quality models.

Measuring quality in use is challenging due to the complexity of current standard models and the incompleteness of other related customized models. The viewpoint of software users is hard to implement within the software life cycle ahead of time, especially for hedonic tasks. Sentiment analysis is proposed to cope with these challenges.

### ACKNOWLEDGMENT

This study was supported in part by Universiti Malaysia Sarawak Zamalah Graduate Scholarship and grant from ERGS/ICT07 (01) /1018/2013 (15).

### REFERENCES

Ahern, D., A. Clouse, and R. Turner. 2008. *CMMI® distilled: A practical introduction to integrated process improvement*, third edition. Boston, MA: Addison-Wesley Professional.

Alaa, G., M. Menshawi, and M. Saeed. 2013. Design criteria and software metrics for efficient and effective web-enabled mobile applications. In *Proceedings of the 8th International Conference on Computer Engineering Systems (ICCES)*, 295–300. New York, NY: IEEE.

Alepis, E., and M. Virvou. 2014. Mobile versus desktop educational applications. In *Object-Oriented User Interfaces for Personalized Mobile Learning.* New York, NY: Springer.

Alnanih, R., O. Ormandjieva, and T. Radhakrishnan. 2013. A new quality-in-use model for mobile user interfaces. In 2013 *Joint Conference of the 23rd International Workshop on Software Measurement and the 2013 Eighth International Conference on Software Process and Product Measurement (IWSM-MENSURA)*, 165–170. New York, NY: IEEE.

Alnanih, R., O. Ormandjieva, and T. Radhakrishnan. 2014. A new methodol-ogy (CON-INFO) for context-based development of a mobile user interface in healthcare applications. In *Pervasive Health State-of-the-Art and Beyond*, first edition, eds. A. Holzinger, M. Ziefle, and C. Röcker, 317–342. London, England: Springer London.

Al-Qutaish, R. E. 2010. Quality models in software engineering literature: An analytical and comparative study. *Journal of American Science* 6, no. 3:166–175.

Ambler, S., J. Nalbone, and M. Vizdos. 2005. *The enterprise unified process: Extending the rational unified process*. Upper Saddle River, NJ: Prentice Hall Press.

Ardito, C., R. Lanzilotti, M. Sikorski, and I. Garnik. 2014. Can evalua-tion patterns enable end users to evaluate the quality of an e-learning system? An exploratory study. In *Universal Access in Human-Computer Interaction. Universal Access to Information and Knowledge*, 185–196. New York, NY: Springer.

Atoum, I., and C. H. Bong. 2014. A framework to predict software "quality in use" from software reviews. In *Proceedings of the First International Conference on Advanced Data and Information Engineering*, eds. J. Herawan et al., 429–436. Singapore: Springer.

Bačíková, M., and J. Porubän. 2014. Domain usability, user's perception. In *Human-Computer Systems Interaction: Backgrounds and Applications*, 15–26. New York, NY: Springer.

Bakota, T., P. Hegedus, P. Kortvelyesi, R. Ferenc, and T. Gyimothy. 2011. A probabilistic software quality model. *In Proceedings of the 2011 27th IEEE International Conference on Software Maintenance (ICSM)*, 243–252, Williamsburg, VA.

Bansiya, J., and C. G. Davis. 2002. A hierarchical model for object-oriented design quality assessment. *IEEE Transactions on Software Engineering* 28, no. 1:4–17.

Becker, P., P. Lew, and L. Olsina. 2012. Specifying process views for a mea-surement, evaluation, and improvement strategy. *Advances in Software Engineering 2012*, 2:2–2:2.

Bevan, N. 1999. Quality in use: Meeting user needs for quality. *Journal of Systems and Software* 49, no. 1:89–96.

Bevan, N. 2001. Quality in use for all. In *User Interfaces for All, Concepts, Methods and Tools*, 352–368. London, England: Lawrence Erlbaum Publications.

Blei, D. M. D., A. Y. A. Ng, and M. I. Jordan. 2003. Latent Dirichlet alloca-tion. *Journal of Machine Learn. Res.* 3:993–1022.

Blei, D. M., and J. D. McAuliffe. 2010. Supervised topic models. *arXiv Preprint arXiv:1003.0783*.

Boehm, B. W., J. R. Brown, H. Kaspar, M. Lipow, G. J. MacLeod, and M. J. Merrit. 1978. *Characteristics of software quality* 1. Amsterdam, Netherlands: North-Holland Publishing Company.

Bond, R. R., D. D. Finlay, C. D. Nugent, G. Moore, and D. Guldenring. 2014. A usability evaluation of medical software at an expert conference setting. *Computer Methods and Programs in Biomedicine* 113, no. 1:383–395.

Brody, S., and N. Elhadad. 2010. An unsupervised aspect-sentiment model for online reviews. In *Human Language Technologies: The 2010 Annual Conference of the North American Chapter of the Association*

# Measuring Software Quality in Use: State-of-the-Art and Research Challenges

for *Computational Linguistics*, 804–812. Stroudsburg, PA: Association for Computational Linguistics. Available at: http://dl.acm.org/citation.cfm?id=1857999.1858121.

Brooks, P., and B. Hestnes. 2010. User measures of quality of experience: Why being objective and quantitative is important. *IEEE Network* 24, no. 2:8–13.

Carvajal, L., A. M. Moreno, M.-I. Sanchez-Segura, and A. Seffah. 2013. Usability through software design. *IEEE Transactions on Software Engineering* 39, no. 11:1582–1596.

Deissenboeck, F., E. Juergens, K. Lochmann, and S. Wagner. 2009. Software quality models: Purposes, usage scenarios and requirements. In *WOSQ '09. ICSE Workshop on Software Quality*, 9–14.

Deissenboeck, F., S. Wagner, M. Pizka, S. Teuchert, and J. F. Girard. 2007. An activity-based quality model for maintainability. In *ICSM 2007 IEEE International Conference on Software Maintenance*, 184–193.

Deming, W. E. 2000. *Out of the crisis*. Cambridge, MA: MIT Press.

Dromey, R. G. 1995. A model for software product quality. *IEEE Transactions on Software Engineering* 21, no. 2:146–162.

Duric, A., and F. Song. 2012. Feature selection for sentiment analysis based on content and syntax models. *Decision Support Systems* 53, no. 4:704–711.

Fogg, B. J., J. Marshall, O. Laraki, A. Osipovich, C. Varma, N. Fanget al. 2001. What makes web sites credible? A report on a large quantitative study. In *Proceedings of the SIGCHI Conference on Human Factors in Computing Systems*, 61–68.

Ganu, G., N. Elhadad, and A. Marian. 2009. Beyond the stars: Improving rating predictions using review text content. In *12th International Workshop on the Web and Databases*.

García, F., M. F. Bertoa, C. Calero, A. Vallecillo, F. Ruíz, M. Piattini, and M. Genero. 2006. Towards a consistent terminology for software measure-ment. *Information and Software Technology* 48, no. 8:631–644.

Garvin, D. A. 1984. What does product quality really mean. *Sloan Management Review* 26, no. 1:25–43.

Ghose, A., and P. G. Ipeirotis. 2011. Estimating the helpfulness and economic impact of product reviews: Mining text and reviewer characteristics. *IEEE Transactions on Knowledge and Data Engineering* 23, no. 10:1498–1512.

González, J. L., R. García, J. M. Brunetti, R. Gil, and J. M. Gimeno. 2012. SWET-QUM: A quality in use extension model for semantic web explo-ration tools. In *Proceedings of the 13th International Conference on Interacci\ón Persona-Ordenador* 15:1–15:8. New York, NY: ACM.

Grady, R. B. 1992. *Practical software metrics for project management and process improvement*. Upper Saddle River, NJ: Prentice-Hall, Inc.

Heinrich, R. 2014. Business process quality. In *Aligning Business Processes and Information Systems* XXII. Wiesbaden, Germany: Springer Fachmedien Wiesbaden.

Heinrich, R., A. Kappe, and B. Paech. 2011. Modeling quality information within business process models. In *Proceedings of the 4th SQMB Workshop, TUM-I1104*, 4–13. Available at: http://se.ifi.uni-heidelberg.de/fileadmin/pdf/Heinrich_ModelingQuality_2011.pdf.

Hirasawa, N. 2013. Challenges for incorporating "quality in use" in embed-ded system development. In *Human Interface and the Management of Information. Information and Interaction for Learning, Culture, Collaboration and Business*, 467–474. New York, NY: Springer.

Holzinger, A. 2005. Usability engineering methods for software developers. *Communications of the ACM* 48, no. 1:71–74.

Horkoff, J., F. B. Aydemir, F.-L. Li, T. Li, and J. Mylopoulos. 2014. Evaluating modeling languages: An example from the requirements domain. In *Conceptual Modeling*, 260–274. New York, NY: Springer.

Hsu, C.-C., and Y.-S. Lee. 2014. Exploring the critical factors influenc-ing the quality of blog interfaces using the decision-making trial and evaluation laboratory (DEMATEL) method. *Behaviour and Information Technology* 33, no. 2:184–194.

Hu, M., and B. Liu. 2004. Mining and summarizing customer reviews. In *Proceedings of the 10th ACM SIGKDD International Conference on Knowledge Discovery and Data Mining*, 168–177. New York, NY: ACM.

Huang, P., and M. Chiu. 2014. Evaluating the healthcare management system by usability testing. In *Digital Human Modeling. Applications in Health, Safety, Ergonomics and Risk Management* 8529, 369–376. New York, NY: Springer. Available at: http://link.springer.com/chapter/10.1007/978-3-319-07725-3_37.

Huang, Z., and M. Benyoucef. 2014. Usability and credibility of e-gov-ernment websites. *Government Information Quarterly* 31, no. 4:584–595.

Ishikawa, K. 1985. *What is total quality control the Japanese way*. Upper Saddle River, NJ: Prentice Hall.

ISO. 1998. ISO 9241-11. *Ergonomic requirements for office work with visual display terminals (VDTs)—Part II guidance on usability*. Geneva, Switzerland: International Organization for Standardization.

ISO/IEC. 2005. ISO/IEC 25000:2005, Software and system engineering—Software product Quality Requirements and Evaluation (SQuaRE)—Guide to SQuaRE. *International Organization for Standarization*. Available at: http://www.iso.org/iso/catalogue_detail.htm?csnumber=35683.

ISO/IEC. 2011. ISO/IEC 25010: 2011, Systems and software engineer-ing—Systems and software quality requirements and evaluation (SQuaRE)—System and software quality models. Geneva, Switzerland: International Organization for Standardization. Available at: http://www.iso.org/iso/home/store/catalogue_tc/catalogue_detail.htm?csnumber=35733.

Jacobson, I. 1999. *The unified software development process*. India: Pearson Education.

Kitchenham, B., S. Linkman, A. Pasquini, and V. Nanni. 1997. The SQUID approach to defining a quality model. *Software Quality Journal* 6, no. 3:211–233.

Kläs, M., J. Heidrich, J. Münch, and A. Trendowicz. 2014. Comprehensive landscapes for software-related quality models. In *Proceedings of the Workshop "Software-Qualit"atsmodellierung und -bewertung* 1403. Kaiserslautern, Germany: Arvix Priprints. Available at: http://arxiv.org/abs/1403.5432.

Kruchten, P. 2004. *The rational unified process: An introduction*. Boston: Addison-Wesley Professional.

La, H. H. J., and S. D. S. Kim. 2013. A model of quality-in-use for service-based mobile ecosystem. In *2013 1st International Workshop on the Engineering of Mobile-Enabled Systems (MOBS)*, 13–18. New York, NY: IEEE.

Leopairote, W., A. Surarerks, and N. Prompoon. 2012. Software quality in use characteristic mining from customer reviews. In *2012 Second International Conference on Digital Information and Communication Technology and its Applications (DICTAP)*, 434–439. New York, NY: IEEE.

Lew, P., L. Olsina, and L. Zhang. 2010. Quality, quality in use, actual usabil-ity and user experience as key drivers for web application evaluation. In *Proceedings of the 10th international conference on Web engineering*, 218–232, Vienna, Austria.

Long, C., J. Zhang, and X. Zhut. 2010. A review selection approach for accurate feature rating estimation. In *Proceedings of the 23rd International Conference on Computational Linguistics: Posters*, 766–774. Beijing, China.

Lu, Y., C. Zhai, and N. Sundaresan. 2009. Rated aspect summarization of short comments. In *Proceedings of the 18th International Conference on World Wide Web*, 131–140. New York, NY: ACM.

Marinescu, R., and D. Ratiu. 2004. Quantifying the quality of object-oriented design: The factor-strategy model. In *Proceedings of the 11th Working Conference on Reverse Engineering*, 192–201, Delft, Netherlands.

Masip, L., M. Oliva, and T. Granollers. 2011. User experience specification through quality attributes. In *Proceedings of the 13th IFIP TC 13 International Conference on Human-Computer Interaction - Volume Part IV*, 656–660. Berlin, Heidelberg: Springer-Verlag. Available at: http://link.springer.com/chapter/10.1007/978-3-642-23768-3_106.

McCall, J. A., P. K. Richards, and G. F. Walters. 1977. *Factors in software quality*. Springfield, VA: General Electric, National Technical Infor-mation- Service.

Mei, Q., X. Ling, M. Wondra, H. Su, and C. Zhai. 2007. Topic sentiment mixture: Modeling facets and opinions in weblogs. In *Proceedings of the 16th International Conference on World Wide Web*, 171–180. New York, NY: ACM.

Mordal, K., N. Anquetil, J. Laval, A. Serebrenik, B. Vasilescu, and S. Ducasse. 2013. Software quality metrics aggregation in industry. *Journal of Software: Evolution and Process* 25, no. 10:1117–1135.

Nielsen, J. 1999. *Designing web usability: The practice of simplicity*. Thousand Oaks, CA: New Riders Publishing.

Oliveira, J., A. Tereso, and R. J. Machado. 2014. An application to select collaborative project management software tools. In *New Perspectives in Information Systems and Technologies, Volume 1*, 467–476. New York, NY: Springer.

Orehovački, T. 2011. Development of a methodology for evaluating the quality in use of web 2.0 applications. In *Human-Computer Interaction— INTERACT 2011*, 382–385. New York, NY: Springer.

Orehovački, T., A. Granić, and D. Kermek. 2013. Evaluating the perceived and estimated quality in use of Web 2.0 applications. *Journal of Systems and Software* 86, no. 12:3039–3059.

Osman, N. B., and I. M. Osman. 2013. Attributes for the quality in use of mobile government systems. In *2013 International Conference on Computing, Electrical and Electronics Engineering (ICCEEE)*, 274–279.

Qiu, G., B. Liu, J. Bu, and C. Chen. 2009. Expanding domain sentiment lexi-con through double propagation. In *Proceedings of the 21st International Joint Conference on Artifical Intelligence*, 1199–1204.

Samadhiya, D., S.-H. Wang, and D. Chen. 2010. Quality models: Role and value in software engineering. In *2010 2nd International Conference on Software Technology and Engineering (ICSTE)*, vol. 1:V1-320 –V1-324.

Scholtz, J. 2006. Beyond usability: Evaluation aspects of visual analytic environments. In *2006 IEEE Symposium on Visual Analytics Science and Technology*, 145–150.

Shackel, B., and S. J. Richardson. 1991. *Human factors for informatics usability*. New York, NY: Cambridge University Press.

Shein, K. P. P., and T. T. S. Nyunt. 2010. Sentiment classification based on ontology and SVM classifier. *2010 Second International Conference on Communication Software and Networks*, 169–172, Washington, DC.

Shewhart, W. A. 1931. *Economic control of quality of manufactured prod-uct*. Van Nostrand Company, Inc.

Stelzer, D., W. Mellis, and G. Herzwurm. 1997. A critical look at ISO 9000 for software quality management. *Software Quality Journal* 6, no. 2:65–79.

Taboada, M., J. Brooke, M. Tofiloski, K. Voll, and M. Stede. 2011. Lexicon-based methods for sentiment analysis. *Comput. Linguist* 37, no. 2:267–307.

Thet, T. T., J.-C. Na, and C. S. Khoo. 2010. Aspect-based sentiment analysis of movie reviews on discussion boards. *Journal of Information Science*.

Turney, P. D. P. 2002. Thumbs up or thumbs down? Semantic orientation applied to unsupervised classification of reviews. In *Proceedings of the 40th Annual Meeting on Association for Computational Linguistics*, 417–424. Stroudsburg, PA: Association for Computational Linguistics.

Wagner, S., K. Lochmann, L. Heinemann, M. Kläs, A. Trendowicz, R. Plösch, and J. Streit. 2012. The quamoco product quality modelling and assess-ment approach. In *Proceedings of the 2012 International Conference on Software Engineering*, 1133–1142. Piscataway, NJ: IEEE Press. Available at: http://dl.acm.org/citation.cfm?id=2337223.2337372.

Weinerth, K., V. Koenig, M. Brunner, and R. Martin. 2014. Concept maps: A useful and usable tool for computer-based knowledge assessment? A literature review with a focus on usability. *Computers & Education* 78, no. 0:201–209.

Zhang, L., and B. Liu. 2011. Identifying noun product features that imply opinions. In *Proceedings of the 49th Annual Meeting of the Association for Computational Linguistics: Human Language Technologies: Short papers*, 2:575–580.

Zhang, L., B. Liu, S. S. H. Lim, and E. O'Brien-Strain. 2010. Extracting and ranking product features in opinion documents. In *Proceedings of the 23rd International Conference on Computational Linguistics: Posters*, 1462–1470. Stroudsburg, PA: Association for Computational Linguistics. Available at: http://dl.acm.org/citation.cfm?id=1944566.1944733.

Zhang, W., H. Xu, and W. Wan. 2012. Weakness finder: Find product weak-ness from Chinese reviews by using aspects based sentiment analysis. E*xpert Systems with Applications* 39, no. 11:10283–10291.

Zheng, X., P. Martin, and K. Brohman, L. Da Xu. 2014. CLOUDQUAL: A qual-ity model for cloud services. *IEEE Transactions on Industrial Informatics* 10, no. 2:1527–1536.

## BIOGRAPHIES

**Issa Atoum** is a doctoral student at the Universiti Malaysia Sarawak, Malaysia. His current research is on sentiment analysis. He has more than 15 years of professional experience in IT services, project management, and quality assurance. He is a certified Project Management Professional, ITIL®V3, and ISO/IEC 20000. He worked as a project manager for the government of Abu Dhabi and Dubai, UAE for mid- and large-scale projects in the domain of security and IT services. His major areas of interest are cyber security, software quality, and sentiment analysis. Atoum can be reached by email at issa.atoum@gmail.com.

**Bong Chih How** is an assistant professor in the Department of Computer Science and Information Technology, Universiti Malaysia Sarawak, Malaysia. His main interests are to use computer technology to solve real-life problems. His current research focuses on natural language processing and information retrieval and their application to pressing problems in information system, psychometrics, and discourse understanding.